

\input phyzzx.tex
\overfullrule0pt

\def\ie{{\it i.e.}}
\def\etal{{\it et al.}}
\def\half{{\textstyle{1 \over 2}}}

\def\bold#1{\setbox0=\hbox{$#1$}%
     \kern-.025em\copy0\kern-\wd0
     \kern.05em\copy0\kern-\wd0
     \kern-.025em\raise.0433em\box0 }
\Pubnum={VAND-TH-94-7-R}
\date={April 1994}
\pubtype{}
\titlepage


\vskip1cm
\title{\bf Closing the Light Gluino Window in Supersymmetric Grand
Unified Models}
\author{Marco Aurelio D\'\i az }
\vskip .1in
\centerline{Department of Physics and Astronomy}
\centerline{Vanderbilt University, Nashville, TN 37235}
\vskip .2in

\centerline{\bf Abstract}
\vskip .1in

We study the light gluino scenario giving special attention to
constraints from the masses of the
light CP-even neutral Higgs $m_h$, the lightest chargino
$m_{\chi^{\pm}_1}$, and the second lightest neutralino
$m_{\chi^0_2}$, and from the $b\rightarrow s\gamma$ decay.
We find that minimal $N=1$ supergravity, with a radiatively broken
electroweak symmetry group and universality of scalar and gaugino
masses at the unification scale, is incompatible with
the existence of a light gluino.

\vskip 1cm
\centerline{Revised Version}

\vfill

\endpage

\voffset=-0.2cm

\REF\lightold{G.R. Farrar and P. Fayet, {\it Phys. Lett.} {\bf
76B}, 575 (1978); T. Goldman, {\it Phys. Lett.} {\bf 78B}, 110
(1978); I. Antoniadis \etal, {\it Nucl.
Phys.} {\bf B211}, 216 (1983); M.J. Eides and M.J. Vysotsky,
{\it Phys. Lett.} {\bf 124B}, 83 (1983); E. Franco, {\it Phys.
Lett.} {\bf 124B}, 271 (1983); G.R. Farrar, {\it Phys. Rev.
Lett.} {\bf 53}, 1029 (1984);
J. Ellis and H. Kowalski, {\it Phys. Lett. B} {\bf
157}, 437 (1985); V. Barger \etal,
{\it Phys. Rev. D} {\bf 33}, 57 (1986).}
\REF\lightnew{I. Antoniadis \etal, {\it
Phys. Lett. B} {\bf 262}, 109 (1991); L. Clavelli, {\it Phys.
Rev. D} {\bf 46}, 2112 (1992); M. Jezabek and J.H. K\"uhn, {\it
Phys. Lett. B} {\bf 301}, 121 (1993); J. Ellis \etal,
{\it Phys. Lett. B} {\bf 305}, 375 (1993);
R.G. Roberts and W.J. Stirling, {\it Phys. Lett. B} { \bf 313},
453 (1993); M. Carena \etal,
{\it Phys. Lett. B} {\bf 317}, 346 (1993);
F. Cuypers, {\it Phys. Rev. D} {\bf 49}, 3075 (1994);
R. Mu\~noz-Tapia and W.J. Stirling, Report No. DTP/93/72, Sept.
1993; D.V. Shirkov and S.V. Mikhailov, Report No. BI-TP 93/75,
Jan. 1994.}
\REF\CValle{F. de Campos and J.W.F. Valle, Report No. FTUV/93-9,
Feb. 1993.}
\REF\LNW{J.L. Lopez \etal, {\it Phys.
Lett. B} {\bf 313}, 241 (1993).}

Discussions about the existence of a light gluino
have been in the literature for a long time
\refmark\lightold. Recently, motivated by the discrepancy between the
value of the strong coupling
constant determined by low energy deep inelastic
lepton-nucleon scattering, and the one
determined by high energy $e^+e^-$ LEP experiments,
there has been a renewed interest in this
possibility\refmark{\lightnew-\LNW}.

\REF\howie{H.E. Haber, Report No. SCIPP 93/21, July 1993.}
\REF\BHaberK{R.M. Barnett \etal, {\it Nucl.
Phys.} {\bf B267}, 625 (1986).}

As it was explained by H.E. Haber\refmark\howie, the light gluino
window is:
$$2.6\lsim m_{\tilde g}\lsim 3\,{\rm GeV},\eqn\mgluino$$
where the lower limit comes from the non-observation of a pseudoscalar
$\tilde g\tilde g$ bound state in quarkonium decays, and the upper
limit follows from an analysis of CERN $p\bar p$
Collider data\refmark\BHaberK.

\REF\susyrep{H.P. Nilles, {\it Phys. Rep.} {\bf 110}, 1 (1984);
H.E. Haber and G.L. Kane, {\it Phys. Rep.} {\bf 117}, 75 (1985);
R. Barbieri, {\it Riv. Nuovo Cimento} {\bf 11}, 1 (1988).}

In supersymmetric grand unified theories
(SUSY-GUT)\refmark\susyrep the three
gaugino masses $M_s$, $M$, and $M'$ are different at the weak scale
but equal to a common gaugino mass $M_{1/2}$ at the grand unification
scale $M_X$. The difference at the weak scale is due to the fact that
the evolution of the three masses is controlled by different
renormalization group equations (RGE). The approximated solution of
these RGE is:
$$\eqalign{M_s&\approx M_{1/2}\left[1+{{3g_s^2}\over
{8\pi^2}}\ln{{M_X}\over{m_Z}}\right],\qquad m_{\tilde g}=|M_s|,
\cr M&\approx M_{1/2}\left[1-{{g^2}\over{8\pi^2}}\ln{{M_X}\over{m_Z}}
\right],
\cr M'&\approx M_{1/2}\left[1-{{11{g'}^2}\over{8\pi^2}}\ln{{M_X}\over{
m_Z}}\right],\cr}\eqn\RGEsol$$
where we are neglecting the supersymmetric
threshold effects. Taking $M_X=10^{16}$ GeV, we find that $M\approx
0.30\,m_{\tilde g}$ and $M'\approx 0.16\,m_{\tilde g}$ at the weak
scale.

Similarly,
the scalar masses are also degenerate at the unification scale, and
equal to $m_0$. The RGE make both the Higgs mass parameters
$m_1$ and $m_2$, and the squark and slepton mass parameters,
evolve differently.
A third independent parameter at the unification scale
is the mass parameter $B$. This mass defines the value of the unified
trilinear mass parameter $A$ at $M_X$ by $A=B+m_0$, a
relation valid in models with canonical kinetic terms. Moreover, it
also defines the
third Higgs mass parameter $m_{12}^2=-B\mu$, valid at every scale,
where $\mu$ is the supersymmetric Higgs mass parameter.
The set of independent parameters we choose to work with, given by
$M_{1/2}$, $m_0$, and $B$ at the unification scale,
is completed by the value of the top quark Yukawa coupling
$h_t=gm_t/(\sqrt{2}m_Ws_{\beta})$
at the weak scale. Here the angle $\beta$ is defined through
$\tan\beta=v_2/v_1$, where $v_1$ and $v_2$ are the vacuum expectation
values of the two Higgs doublets.

\REF\neutralinorad{A.B. Lahanas \etal,
{\it Phys. Lett. B} {\bf 324}, 387 (1994);
D. Pierce and A. Papadopoulos, Report No.
JHU-TIPAC-930030, Dec. 1993, and JHU-TIPAC-940001,
Feb. 1994.}

Knowing the parameters of the Higgs potential at the weak scale
$m_1^2$, $m_2^2$, and $B$, we can calculate the more familiar
parameters $m_t$, $t_{\beta}$, $m_A$, and $\mu$, for a given
value of the top quark Yukawa coupling $h_t$, through the
following formulas
$$\eqalign{m_{1H}^2\equiv m_1^2+\mu^2=&-\half m_Z^2c_{2\beta}+\half
m_A^2(1-c_{2\beta}),
\cr m_{2H}^2\equiv m_2^2+\mu^2=&\half m_Z^2c_{2\beta}+\half
m_A^2(1+c_{2\beta}),\cr
m_{12}^2=-B\mu=&\half m_A^2s_{2\beta},\cr}\eqn\conversion$$
where $s_{2\beta}$ and $c_{2\beta}$ are sine and cosine functions of
the angle $2\beta$, and it is understood that all the parameters
are evaluated at the weak scale. We alert the reader that for a
given set of values $M_{1/2}$, $m_0$, $B$, and $h_t$ there may
exist more than one solution for the parameters at the weak scale
$m_t$, $t_{\beta}$, $m_A$, and $\mu$. According to ref.~[\LNW],
and we will confirm this, the relevant region of parameter space
in the light gluino scenario is characterized by low values of the
top quark mass and values of $\tan\beta$ close to unity.
Considering the low values of the top
quark mass relevant for our calculations,
radiative corrections to the chargino and
neutralino masses (recently calculated in ref.~[\neutralinorad])
will have a minor effect.

\REF\mbmtau{M. Carena \etal, {\it Nucl.
Phys.} {\bf B406}, 59 (1993); P. Langacker and N. Polonsky,
{\it Phys. Rev. D} {\bf 49}, 1454 (1994).}
\REF\AnantBS{B. Ananthanarayan \etal, Report
No. BA-94-02, Jan. 1994.}
\REF\diazhaberii{M.A. D\'\i az and H.E. Haber, {\it Phys. Rev. D}
{\bf 46}, 3086 (1992).}
\REF\aleph{D. Buskulic \etal, {\it Phys. Lett. B} {\bf 313}, 312
(1993).}

The region $\tan\beta$ close to unity has been singled out by the
grand unification condition $m_b=m_{\tau}$ at $M_X$\refmark\mbmtau,
and was analyzed in detail in ref.~[\AnantBS]. Here we stress the
fact that if $\tan\beta=1$, the lightest CP-even neutral Higgs is
massless at tree level. Nevertheless, the supersymmetric
Coleman-Weinberg mechanism\refmark\diazhaberii generates a mass $m_h$
different from zero via radiative corrections. The fact that
$m_t$ is also small will result in a radiatively generated $m_h$
close to the experimental lower limit $m_h\gsim56$ GeV, valid for
$m_A>100$ GeV\refmark\aleph. Therefore, experimental lower limits
on $m_h$ impose important restrictions on the light gluino window.

\REF\BBMR{S. Bertolini \etal,
{\it Nucl. Phys.} {\bf B353}, 591 (1991).}
\REF\masSUSY{M.A. D\'\i az, {\it Phys. Lett. B} {\bf 304},
278 (1993); N. Oshimo, {\it Nucl. Phys.} {\bf B404}, 20 (1993);
J.L. Lopez \etal,
{\it Phys. Rev. D} {\bf 48}, 974 (1993);
Y. Okada, {\it Phys. Lett. B} {\bf 315}, 119 (1993);
R. Garisto and J.N. Ng, {\it Phys. Lett. B} {\bf 315}, 372 (1993);
J.L. Lopez \etal,
{\it Phys. Rev. D} {\bf 49}, 355 (1994);
M.A. D\'\i az, {\it Phys. Lett. B} {\bf 322}, 207 (1994);
F.M. Borzumati, Report No. DESY 93-090, August 1993; S. Bertolini
and F. Vissani, Report No. SISSA 40/94/EP, March 1994.}
\REF\misiak{M. Misiak, {\it Nucl. Phys.} {\bf B393}, 23 (1993).}
\REF\ChargedH{J.F. Gunion and A. Turski, {\it Phys. Rev. D} {\bf 39},
2701 (1989); {\bf 40}, 2333 (1989); A. Brignole \etal,
{\it Phys. Lett.
B} {\bf 271}, 123 (1991); M. Drees
and M.M. Nojiri, {\it Phys. Rev. D} {\bf 45}, 2482 (1992);
A. Brignole, {\it Phys. Lett. B} {\bf 277}, 313 (1992);
P.H. Chankowski \etal,
{\it Phys. Lett. B} {\bf 274}, 191 (1992);
M.A. D\'\i az and H.E. Haber, {\it Phys. Rev. D}
{\bf 45}, 4246 (1992).}
\REF\diaz{M.A. D\'\i az, {\it Phys. Rev. D} {\bf 48}, 2152 (1993).}

It has been pointed out that the branching ratio
$B(b\longrightarrow s\gamma)$ has a strong dependence on the
supersymmetric parameters\refmark{\BBMR,\masSUSY}.
The theoretical branching ratio must remain within the
experimental bounds $0.65\times10^{-4}<B(b\longrightarrow s\gamma)<
5.4\times10^{-4}$. We calculate this ratio, including loops involving
$W^{\pm}$/U-quarks, $H^{\pm}$/U-quarks, $\chi^{\pm}$/U-squarks,
and $\tilde g$/D-squarks, neglecting only the contribution from
the neutralinos, which were
reported to be small\refmark\BBMR. We also include
QCD corrections to the branching ratio\refmark\misiak and
one loop electroweak corrections to both the charged Higgs
mass\refmark\ChargedH and the charged Higgs-fermion-fermion
vertex\refmark\diaz.

\REF\neuDelphi{G. Wormser, published in {\it Dallas HEP 1992},
page 1309.}
\REF\Decamprep{D. Decamp \etal (ALEPH Collaboration), {\it Phys. Rep.}
{\bf 216}, 253 (1992).}

Another important source of constraints comes from the
chargino/neutralino sector. For $\tan\beta\gsim4$, a neutralino with
mass lower than 27 GeV is excluded, but the lower bound decreases
when $\tan\beta$ decreases, and no bound is obtained if
$\tan\beta<1.6$\refmark\neuDelphi. The lower bound for the heavier
neutralinos (collectively denoted by $\chi'$) is $m_{\chi'}>45$ GeV
for $\tan\beta>3$, and this bound also decreases with $\tan\beta$
and eventually disappears\refmark\Decamprep.
On the other hand, if the
lightest neutralino has a mass $\lsim 40$ GeV (as we will see,
in the light gluino scenario, the lightest neutralino has a mass
of the order of 1 GeV), the lower bound
for the lightest chargino mass is 47 GeV\refmark\Decamprep. For
notational convenience, this latest experimental bound will be denoted
by $\bar m_{\chi^{\pm}_1}\equiv47 {\rm GeV}$.

\REF\BHaber{R.M. Barnett and H.E. Haber, {\it Phys. Rev. D} {\bf
31}, 85 (1985).}

In the following, we study the chargino/neutralino sector in
more detail by analysing the mass matrices.
The chargino mass matrix\refmark\BHaber has
eigenvalues denoted by $m_{\chi^{\pm}_i}$, $i=1,2$ and
$m_{\chi^{\pm}_1}<m_{\chi^{\pm}_2}$.
In the light gluino case we have $M\ll m_W$, and the chargino masses
can be approximated by
$$m_{\chi^{\pm}_{1,2}}^2=\half\mu^2+m_W^2\pm\half\sqrt{R}
\pm{{2m_W^2\mu Ms_{2\beta}}\over{\sqrt{R}}}
+{\cal O}(M_{1/2}),\eqn\chaappro$$
where $R=\mu^4+4m_W^2\mu^2+4m_W^4c_{2\beta}^2$.
Since the lightest chargino mass is bounded from below,
we get the following constraint:
$$m_W^4c_{2\beta}^2+\mu^2\bar m_{\chi^{\pm}_1}^2<
\left(m_W^2-\bar m_{\chi^{\pm}_1}^2\right)^2-{{4m_W^2\mu
Ms_{2\beta}(\half\mu^2+m_W^2-\bar m_{\chi^{\pm}_1}^2)}\over{\sqrt{
\mu^2+4m_W^2\mu^2+4m_W^4c_{2\beta}^2}}},
\eqn\chaconst$$
plus terms of ${\cal O}(M_{1/2}^2)$. This limits the values of
$\mu$ and $\tan\beta$:
$$\eqalign{
\mu^2&<\bar m_{\chi^{\pm}_1}^2\left({{m_W^2}\over{\bar
m_{\chi^{\pm}_1}^2}}-1\right)^2-{{4m_W^2\mu_0M(\half\mu_0^2+m_W^2-
\bar m_{\chi^{\pm}_1}^2)}\over
{\bar m_{\chi^{\pm}_1}^2|\mu_0|\sqrt{\mu_0^2+4m_W^2}}}
+{\cal O}(M_{1/2}^2)\cr
&\qquad\qquad\Longrightarrow|\mu|\lsim(90\mp0.87m_{\tilde g})\,{\rm GeV},
\,\,{\rm with}\qquad \pm={\rm sign}(\mu M),\cr
|c_{2\beta}|&<1-{{\bar m_{\chi^{\pm}_1}^2}\over{m_W^2}}+{\cal O}
(M_{1/2}^2)\quad\Longrightarrow\quad
0.46<t_{\beta}<2.2\,,\cr}\eqn\constrain$$
where $\mu_0^2=\bar m^2_{\chi^{\pm}_1}(m_W^2/\bar m_{\chi^{\pm}_1}^2
-1)^2\approx90$ GeV is the zeroth order solution ($M=0$), and
$\mp0.87m_{\tilde g}$ correspond to the first order correction.
The type of constraints given
in eq.~\constrain\ were already found in
ref.~[\LNW] at zeroth order, but as we will see, the neutralino sector
will restrict the parameter space even more.

\REF\GunionHaber{J.F. Gunion, H.E. Haber, {\it Nucl. Phys.} {\bf B272},
1 (1986).}

The neutralino mass matrix\refmark\GunionHaber in the zero
gluino mass limit ($M=M'=0$) has one eigenvalue equal to zero.
Calculating the first order correction, we find for the lightest
neutralino mass:
$$m_{\chi^0_1}=Ms_W^2+M'c_W^2+{\cal O}(M^2_{1/2})\approx
0.19m_{\tilde g},\eqn\lightneut$$
where we used the relations between $M$, $M'$, and $m_{\tilde g}$
given below eq.~\RGEsol. Considering eq.~\mgluino\ we get
$$0.49\lsim\,m_{\chi^0_1}\lsim0.57\,{\rm GeV}.\eqn\neuuno$$
This light neutralino (the lightest supersymmetric particle, or LSP)
is, up to terms of ${\cal O}(M_{1/2}^2/m_Z^2)$, almost a pure photino,
and there is no bound on its mass from LEP collider data. Nevertheless,
in the case of a stable LSP (R-parity conserving models), ref.~[\LNW]
pointed out some cosmological implications that make this scenario
less attractive. On the other hand, the possibility of having a
small amount of R-parity violation is not ruled out, in which case
the LSP would not be stable\refmark\CValle.

The other three neutralino masses are, in first approximation,
solutions of the cubic equation
$$m^3_{\chi^0}-(\mu^2+m_Z^2)m_{\chi^0}-s_{2\beta}\mu m_Z^2=0.
\eqn\cubic$$
According to eq.~\constrain, the value of $\tan\beta$ will be close
to unity, \ie, $s_{2\beta}\approx 1$. If we expand around this value
we get for the other neutralino masses:
$$\eqalign{
m_{\chi^0_2}=&-\mu-\mu{{m_Z^2(1-s_{2\beta})}\over{2\mu^2-m_Z^2}},\cr
m_{\chi^0_{3,4}}=&m_{\pm}
+{{m_Z^2(\mu+m_{\pm})(Mc_W^2+M's_W^2)}\over{3\mu m_Z^2+
2(\mu^2+m_Z^2)m_{\pm}}}
-{{\mu m_Z^2m_{\pm}(1-s_{2\beta})}\over{3\mu m_Z^2+
2(\mu^2+m_Z^2)m_{\pm}}}\,,
\cr}\eqn\tresneuapp$$
where $m_{\pm}\equiv\half\mu\pm\half\sqrt{\mu^2+4m_Z^2}$ and
we neglect terms of ${\cal O}(1-s_{2\beta})^2$ and
${\cal O}(M_{1/2}^2)$. It is
understood that if an eigenvalue of the neutralino mass matrix
is negative, a simple rotation of the fields will give us a positive
mass. The approximation in eq.~\tresneuapp\
breaks down when $\mu^2\approx\half m_Z^2$ except for $t_{\beta}=1$.

\FIG\figi{Contours of constant value of the lightest chargino and
the second lightest neutralino
masses, for a gluino mass $m_{\tilde g}=3$ GeV. The contour
corresponding to the chargino mass is
defined by the experimental lower bound
$m_{\chi^{\pm}_1}=47$. For $\chi^0_2$ we plot contour of constant mass
from 5 to 45 GeV (dashed lines). The solid line that joins the
crosses represent the $tan\beta$ dependent bound on $m_{\chi^0_2}$.
The ``allowed'' region lies below the two solid lines.
We are considering in this graph experimental restrictions
from the chargino/neutralino searches only.}

Now we turn to the exact numerical calculation of
the chargino and neutralino masses. In Fig.~\figi\ we plot contours
of constant masses in the $\mu-t_{\beta}$ plane. The
curve $m_{\chi^{\pm}_1}=47$ GeV corresponds to
the constraint expressed in eq.~\chaconst. We also plot contours
defined by $m_{\chi^0_2}=5-45$ GeV, and the $\tan\beta$ dependent
experimental bound on $m_{\chi^0_2}$ is represented by the solid line
that joins the crosses. In this way, the ``allowed'' region
(including chargino/neutralino searches only) corresponds to the region
below the two solid lines. For $\mu<0$ the allowed region
is almost an exact reflection. The approximate bounds
for $\mu$ we got in eq.~\constrain\ are confirmed numerically:
$\mu<87.4$ GeV for $m_{\tilde g}=3$ GeV. Nevertheless, the bounds
on $\tan\beta$ come only from the experimental
result $m_{\chi^{\pm}_2}>47$ GeV, and we must include also the
experimental results on $m_{\chi^0_2}$. From Fig.~\figi\ we see
that this bound restricts the model to
$\tan\beta\lsim1.82$, with the equality valid for
$\mu=49.4$ GeV. Since for $\tan\beta\lsim1$
there is no solution for the radiatively broken
electroweak symmetry group, the allowed values
of $\tan\beta$ in the light gluino scenario and with $\mu>0$ are
$$1\lsim\tan\beta\lsim1.82\,.\eqn\tanballow$$
If $\mu<0$, the upper bound is $\tan\beta\lsim 1.85$ with the equality
valid for $\mu=-51.8$ GeV. We go on to analyze the
viability of the ``allowed'' region in Fig.~\figi. We will find that
the region allowed by the $\chi^{\pm}$ and $\chi^0$ analysis is in fact
disallowed by the experimental bound on $m_h$ and $m_t$.

\REF\solRGE{R. Barbieri and G.F. Giudice, {\it Nucl. Phys.}
{\bf B306}, 63 (1988); M. Matsumoto \etal,
{\it Phys. Rev. D} {\bf 46}, 3966 (1992).}
\REF\topbound{S. Abachi \etal (D0 Collaboration), {\it Phys. Rev.
Lett.} {\bf 72}, 2138 (1994).}

In ref.~[\solRGE]\ the RGE are solved for the special case in which
only the top quark Yukawa coupling is different from zero. In the
case of a light gluino ($M_{1/2}\approx0$),
the value of $\mu$ at the weak scale can be approximated
by\refmark\solRGE
$$\half m_Z^2+\mu^2=-m_0^2+{{z-1}\over{z(1-t_{\beta}^{-2})}}
\left[{{3m_0^2}\over2}+{{A^2}\over{2z}}\right],\eqn\solmu$$
with
$$z^{-1}=1-(1+t_{\beta}^{-2})\left({{m_t}\over{193 {\rm GeV}}}
\right)^2.\eqn\zeta$$
As it was reported in ref.~[\LNW], there is a fine-tuning situation
in which we can have $m_0\gg|\mu|$ (producing larger radiative
corrections to $m_h$) and it is obtained when the
coefficient of $m_0^2$ in eq.~\solmu\ is zero. Ref.~[\LNW]\ concluded
that constraints on $m_h$ can be satisfied in a small window
around $\tan\beta=1.88-1.89$ (they did not consider the constraint
on the second lightest neutralino). We will see that if the
relation $A=B+m_0$ holds we do not find this type of solution
($m_0\gg|\mu|$) as opposed to the case in which $A=0$. However, the
later is obtained for a value of the top quark mass below the
value of the experimental lower bound $m_t\ge131$ GeV\refmark\topbound.

We survey the parameter space $m_0$, $B$, $M_{1/2}$, and $h_t$,
looking for the maximum value
of $\tan\beta$ allowed by collider negative searches in the
chargino/neutralino sector, using the SUSY-GUT model described earlier.
We first consider models in which the relation $A=B+m_0$ holds.
We expect maximum $\tan\beta$ to maximize $m_h$. For example,
for the value $h_t=0.87$ and $M_{1/2}=1$ GeV
(essentially fixed by the light gluino mass hypothesis)
we find that $m_0=132.9$
and $B=-225.5$ GeV (at the unification scale) gives us
$\tan\beta=1.82$ and $\mu=49.4$ GeV, \ie, the
critical point with maximum $\tan\beta$ in the upper
corner of the allowed region in Fig.~\figi. The values of other
important parameters at the weak scale are
$m_{\chi^{\pm}_1}\approx47.1$, $m_{\chi^0_2}\approx36.8$,
$m_t=131.1$, $m_A=152.1$, and $m_{\tilde g}=2.75$ GeV. We find a
value for $B(b\longrightarrow s\gamma)=5.35\times10^{-4}$ consistent
with the CLEO bounds. However, the lightest CP-even neutral Higgs
fails to meet the experimental
requirement: we get $m_h=47.7$ GeV, inconsistent with LEP data.

{}From the two fixed parameters, $h_t$ and $M_{1/2}$, the one that could
affect the mass of the CP-even neutral Higgs is the first one; for a
fixed value of $\tan\beta$, a larger value of the top quark Yukawa
coupling will give us a larger $m_t$, and this will increase $m_h$.
However, $h_t$ also enters the RGE for the Higgs mass parameters,
and in order to get the correct electroweak symmetry breaking, a
smaller value of $m_0$ is necessary. This implies smaller squark
masses, which in turn reduce $m_h$ through radiative corrections.
As an example with a larger $h_t$, we
have found that for $h_t=0.97$ and $M_{1/2}=1$ GeV, the critical
point is obtained at $m_0=103.8$ and $B=-132.5$ GeV. As expected, the
value of the top quark mass is larger ($m_t=146.2$ GeV),
but we get smaller values for the squark masses.
The net effect is that now $m_h$ is even smaller, 43.5
GeV, also in conflict with the experimental lower bound. (We caution
the reader that at the small values of $m_t$ and $m_0$ used here,
the contributions to $m_h$ coming from the
Higgs/Gauge-boson/neutralino/chargino are also
important\refmark\diazhaberii; we include these in our analysis.)

We go back to $h_t=0.87$ to analyze the case
$\mu<0$. In this case the critical point, given by $\tan\beta=1.85$
and $\mu=-51.8$ GeV, is obtained for $m_0=71.1$ and
$B=111$ GeV. However the light CP-even Higgs is lighter than before:
$m_h=40.4$ GeV, incompatible with LEP data.

Models in which $A$ and $B$ are independent parameters have one
extra degree of freedom that may help to satisfy the experimental
constraints. According to eq.~\solmu\ the fine tuning $m_0\gg|\mu|$
is obtained for $A=0$. Adopting that value and considering $\mu>0$,
for $h_t=0.87$ and $M_{1/2}=1$
GeV we obtain the critical point for $m_0=151.6$ and $B=-256.8$ GeV,
which implies $m_{\chi^{\pm}_1}=47.1$, $m_{\chi^0_2}=36.8$,
$m_t=131.1$, $m_A=173.4$, and $m_{\tilde g}=2.75$ GeV. However,
we get $B(b\longrightarrow s\gamma)=7.15\times10^{-4}$ and
$m_h=48.8$ GeV, both inconsistent with experimental bounds.

In order to illustrate the fine-tuning we consider $h_t=0.77$ and
$M_{1/2}=1$ GeV. The critical point is obtained for $m_0=930$
and $B=-9576$ GeV. The masses of $\chi^{\pm}_1$, $\chi^0_2$
and $\tilde g$ are the same as before, and we also get $m_A=1059$,
$m_h=64.8$ GeV consistent with LEP bound,
and $B(b\longrightarrow s\gamma)=4.14\times10^{-4}$ consistent
with the CLEO bound, but this time it is the top quark mass
that does not meet the experimental bound: we get $m_t=116.0$ GeV,
incompatible with the $D0$ lower bound of 131 GeV.

If $\mu<0$ no big changes are found. For $h_t=0.87$ the
critical point, defined now
by $\tan\beta=1.85$ and $\mu=-51.8$ GeV, is obtained for $m_0=159.4$
and $B=268$ GeV and, as before, the two quantities inconsistent
with experimental results are $B(b\longrightarrow s\gamma)=
8.42\times10^{-4}$ and $m_h=50.5$ GeV.

Our conclusion is that N=1 Supergravity with a radiatively broken
electroweak symmetry group is incompatible with a light gluino
with a mass of a few GeV. This is valid
in models where the relation $A=B+m_0$ holds as well as
in models where $A$ and $B$ are independent parameters.

{\bf Acknowledgments:} I am thankful to Howard Baer, Manuel Drees,
Tonnis ter Veldhuis, and Thomas J. Weiler for useful conversations.
This work was supported by the U.S. Department of Energy, grant No.
DE-FG05-85ER-40226, and by the Texas National Research Laboratory
(SSC) Commission, award No. RGFY93-303.

\refout
\figout
\end